\documentclass[preprint,aps,prb,showpacs,showkeys,preprintnumbers,amsmath,amssymb,superscriptaddress]{revtex4}

\usepackage{graphicx}% Include figure files
\begin{document}

\title{Electronic structure and transport anisotropy of Bi${_2}$Te${_3}$ and Sb${_2}$Te${_3}$}
\author{B. Yu. Yavorsky}
\email{bogdan.yavorsky@physik.uni-halle.de}
\affiliation{Institut f\"{u}r Physik, Martin-Luther-Universit\"{a}t Halle-Wittenberg, D-06099 Halle, Germany}
\author{N. F. Hinsche}
\affiliation{Institut f\"{u}r Physik, Martin-Luther-Universit\"{a}t Halle-Wittenberg, D-06099 Halle, Germany}
\
\author{I. Mertig}
\affiliation{Institut f\"{u}r Physik, Martin-Luther-Universit\"{a}t Halle-Wittenberg, D-06099 Halle, Germany}
\affiliation{Max-Planck-Institut f\"{u}r Mikrostrukturphysik, Weinberg 2, D-06120 Halle, Germany}
\author{P. Zahn}
\affiliation{Institut f\"{u}r Physik, Martin-Luther-Universit\"{a}t Halle-Wittenberg, D-06099 Halle, Germany}
\date{\today}% It is always \today, today,
\begin{abstract}
On the basis of detailed \textit{ab initio} studies the influence of 
strain on the anisotropy of the transport distribution of the thermoelectrics 
Bi$_2$Te$_3$ and Sb$_2$Te$_3$ was investigated. Both tellurides were studied in their own, 
as well as in their co-partners lattice structure to gain insight
to the electrical transport in epitaxial heterostructures composed of both materials. 
It is shown, that the anisotropy of the transport distribution overestimates
the experimental findings  for Bi$_2$Te$_3$, implying anisotropic scattering effects. 
An increase of the in-plane lattice constant leads to an enhancement of the
transport anisotropy for $p$-doping, whereas the opposite occurs for $n$-doping.
The recent findings and special features of the transport distribution are
discussed in detail in relation to the topology of the band structures.
\end{abstract}

%Use showkeys class option if keyword
\maketitle
\section{Introduction}
Thermoelectric (TE) materials have huge potential for power generation, heat 
pumping, and refrigeration. However, their practical application is
restricted because of the low performance of the TE devices compared to
traditional fossil fuel power generators and compressor-based
refrigerators \cite{DiSalvo}. A high performance TE material has to be a good
electrical conductor as well as a poor thermal conductor and, at the same time,
possess a large Seebeck coefficient\cite{Bell}. Quantitatively the efficiency of
TE devices is  expressed by the dimensionless figure of merit $ZT$,  
\begin{equation}
ZT= {\alpha ^2}\sigma T/(\kappa _L + \kappa _e),
\end{equation}
where $\alpha$, $\sigma$, $T$ are the Seebeck coefficient, electrical
conductivity and absolute temperature, and $\kappa _L$, $\kappa_e$ are phonon
and electron contributions to the total thermal conductivity, respectively.

Bismuth and antimony tellurides, Bi$_2$Te$_3$, Sb$_2$Te$_3$, and the alloys
based on these materials play a significant role for thermoelectric
technology. Already early studies of the (Bi$_x$,Sb$_{2-x}$)Te$_3$ 
compounds in the late 1950s reported $ZT\sim 1$ at room temperature 
\cite{exp_early}, confirmed by further experiments \cite{exp_further}. This
value remains, even to date, the maximal one available at room temperature for bulk
materials. Current progress in nanostructure fabrication, in particular, epitaxial
growth of high-quality superlattices \cite{Rama1}, encourages the possibility of
significant performance improvement of TE devices. Venkatasubramanian {\it et al.}
reported $ZT\sim 2.4$ and $ZT\sim 1.45$ at 300K for $p$-type and $n$-type
Bi$_2$Te$_3$/Sb$_2$Te$_3$ superlattices, respectively\cite{Rama2}. 

These experimental advances motivated extensive theoretical studies of the
electronic structure of the bulk bismuth and antimony telluride aimed to
understand the possible origin of the increased thermoelectic performance in
the multilayered structures. While in the previous years only few {\it ab
  initio} band-structure calculations of the bulk bismuth telluride
\cite{Thomas,Mishra} could be mentioned, in the last decade various aspects of
the electronic structure of both pure and doped bulk Bi$_2$Te$_3$ and
Sb$_2$Te$_3$  as well as their transport properties were discussed in
Ref.[\onlinecite{Larson1,Youn,Scheidemantel,Thonhauser1,Larson2,Thonhauser2,Kim,Larson3,Lee,
Wang,Huang,Park,Eremeev}]. {\it Ab initio} studies of the electronic structure and 
the transport properties of Bi$_2$Te$_3$/Sb$_2$Te$_3$ superlattices were also
reported\cite{Li}.

An explanation of directional anisotropy of the transport properties in
Bi$_2$Te$_3$/Sb$_2$Te$_3$ superlattices could play a crucial role for the
understanding of their increased figure of merit. Venkatasubramanian {\it et
  al.}\cite{Rama2} found a strong dependence of the anisotropy of the carrier
mobility on both the superlattice period and the relative thickness of the
constituents. The enhancement of the electrical conductivity parallel to the
epitaxial growth direction, i.e. the trigonal axis of the rhombohedral lattice
of bismuth and antimony tellurides, together with the possibility to suppress
the lattice thermal conductivity $\kappa _L$ along this direction could
provide the desirable ZT enhancement.

In this study we concentrate on the anisotropy of the transport properties in the bulk Bi$_2$Te$_3$
and Sb$_2$Te$_3$ as a first step on a way of understanding the corresponding properties of the
Bi$_2$Te$_3$/Sb$_2$Te$_3$ superlattices. Since epitaxial growth always implies
lattice distortions we included, as discussed below, the effect of the lattice
relaxation in our study.

\section{Crystal structure}
\begin{figure}
\includegraphics[scale=0.40]{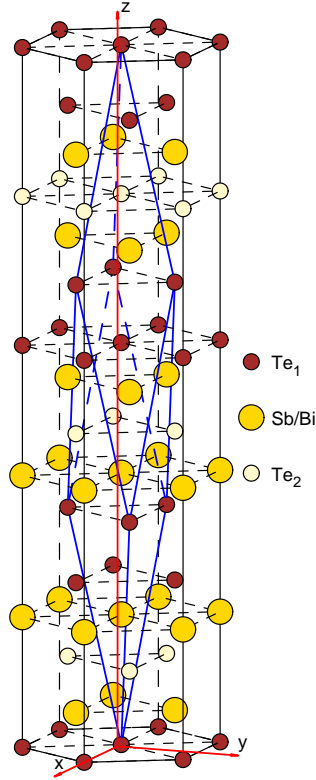}
\caption{\label{structure} The rhombohedral unit cell superimposed with the
  hexagonal one to emphasize the layered character of the material.}
\end{figure}

Both bismuth and antimony telluride possess the rhombohedral crystal structure
with five atoms, i.e. one formula unit, per unit cell belonging to the space
group $D^5_{3d}$ ($R\bar{3}m$). Experimental lattice
parameters\cite{SpringerData} are ${a^{rh}_{BiTe}}=10.473${\AA ,}
$\theta_{BiTe}=24.17^{\circ}$, and ${a^{rh}_{SbTe}}=10.447${\AA ,}
$\theta_{SbTe}=23.55^{\circ}$, were $\theta$ is the angle between the
rhombohedral basis vectors of the length $a^{rh}$. In order to emphasize the
layered character of this structure it is convenient to rearrange it into the
hexagonal unit cell built up by three formula units, as shown in
Fig.\ref{structure}. The hexagonal cell contains 15 atoms grouped in the 3
'quintuple' layers, Te1-Bi(Sb)-Te2-Bi(Sb)-Te1, where Te1 and Te2 are two
different crystal kinds of tellurium atoms. The 'hexagonal'  lattice parameters
are ${a^{hex}_{BiTe}}=4.384${\AA ,} $c^{hex}_{BiTe}=30.487${\AA ,} and
${a^{hex}_{SbTe}}=4.264${\AA ,} $c^{hex}_{SbTe}=30.458${\AA ,} for
Bi$_2$Te$_3$ and Sb$_2$Te$_3$, respectively. In bismuth telluride the
nearest interatomic distances between the individual monolayers inside the
'quintuple' blocks are 3.07{\AA } for Te1-Bi and 3.25{\AA } for Bi-Te2. Two
adjacent 'quintuple' layers in Bi$_2$Te$_3$ are separated by somewhat longer
distance, 3.63{\AA } for Te1-Te1. In the antimony telluride these
distances are 2.98{\AA } for Te1-Sb, 3.17{\AA } for Sb-Te2 inside the
'quintuple' blocks, and 3.74{\AA } for Te1-Te1 between the blocks.

In the Bi$_2$Te$_3$/Sb$_2$Te$_3$ multilayers atoms change their bulk positions 
due to the mismatch of the lattice parameters. The description of the realistic 
crystal structure of the multilayers is out of the scope of this study. 
Nonetheless, we modeled Bi$_2$Te$_3$ with the experimental lattice parameters 
and interatomic distances of Sb$_2$Te$_3$, and vice versa. Since both materials have
very similar lattice parameter $c^{hex}$ along $z$-axis this variation is essentially
compression and extension of the lattice in ($xy$) plane for bismuth and
antimony telluride, respectively. We assume that one could estimate the effect
of the lattice relaxation on the electronic and transport properties in the
Bi$_2$Te$_3$/Sb$_2$Te$_3$ heterostructures from these two limiting cases.

\section{Calculational details}
Calculations of the electronic structures were performed by means of the
screened Korringa-Kohn-Rostoker Green's function method \cite{KKR} in the
atomic sphere approximation (ASA) within the local density approximation of
the density functional theory in the parameterization of Vosko {\it et
  al.}\cite{Vosko}. It is generally recognized that the effects of spin-orbit
coupling are mandatory for the correct treatment of the band structure in
these materials. Therefore we used a fully relativistic version of the method
based on the Dirac equation\cite{KKRrel}. The obtained self-consistent band structures 
were used for the calculations of the transport distribution $\sigma_{\alpha \beta}$
within the Boltzmann formalism assuming a constant relaxation time $\tau$ \cite{Ziman},
\begin{equation}
\sigma_{\alpha \beta}(E)= \tau\frac{e^2}{(2\pi)^3\hbar}\sum_{j}\int\limits_{\varepsilon^j({\bf k})=E} dS
 \frac{v^j_\alpha ({\bf k})v^j_\beta ({\bf k})}{\mid {\bf v}^j ({\bf k})\mid }, \qquad
{\bf v}^j({\bf k})=\frac{1}{\hbar}\bigtriangledown _{\bf k} \varepsilon ^j ({\bf k}),
\end{equation}
where $\varepsilon ^j ({\bf k})$ is the {\it j}-th band energy
at the {\bf k}-point of the Brillouin zone (BZ), $\alpha$ and $\beta$ denote
cartesian coordinates. We assume the relaxation time
to be isotropic. In this approximation the transport anisotropy ratio
$\sigma_{xx}/\sigma_{zz}$ is independent on the relaxation time, and 
we do not have to specify it.

The {\bf k}-space integration over the isoenergetic surfaces was performed
with the tetrahedron method on the Bl{\"o}chl mesh{\cite{Bloechl}} of
96$\times$96$\times$96 points in the whole BZ. In the energy intervals of the
width about 0.1 eV in the vicinity of both band edges the integration was refined
by means of an adaptive mesh up to 4$\times$4$\times$4 times as dense as the
original one. As a convergence criterium we used the values of the transport
anisotropy ratio calculated from the effective mass approximation at the band
extrema \cite{Peter}.

\section{Electronic structures and transport properties}
\begin{figure}
\includegraphics[scale=0.35]{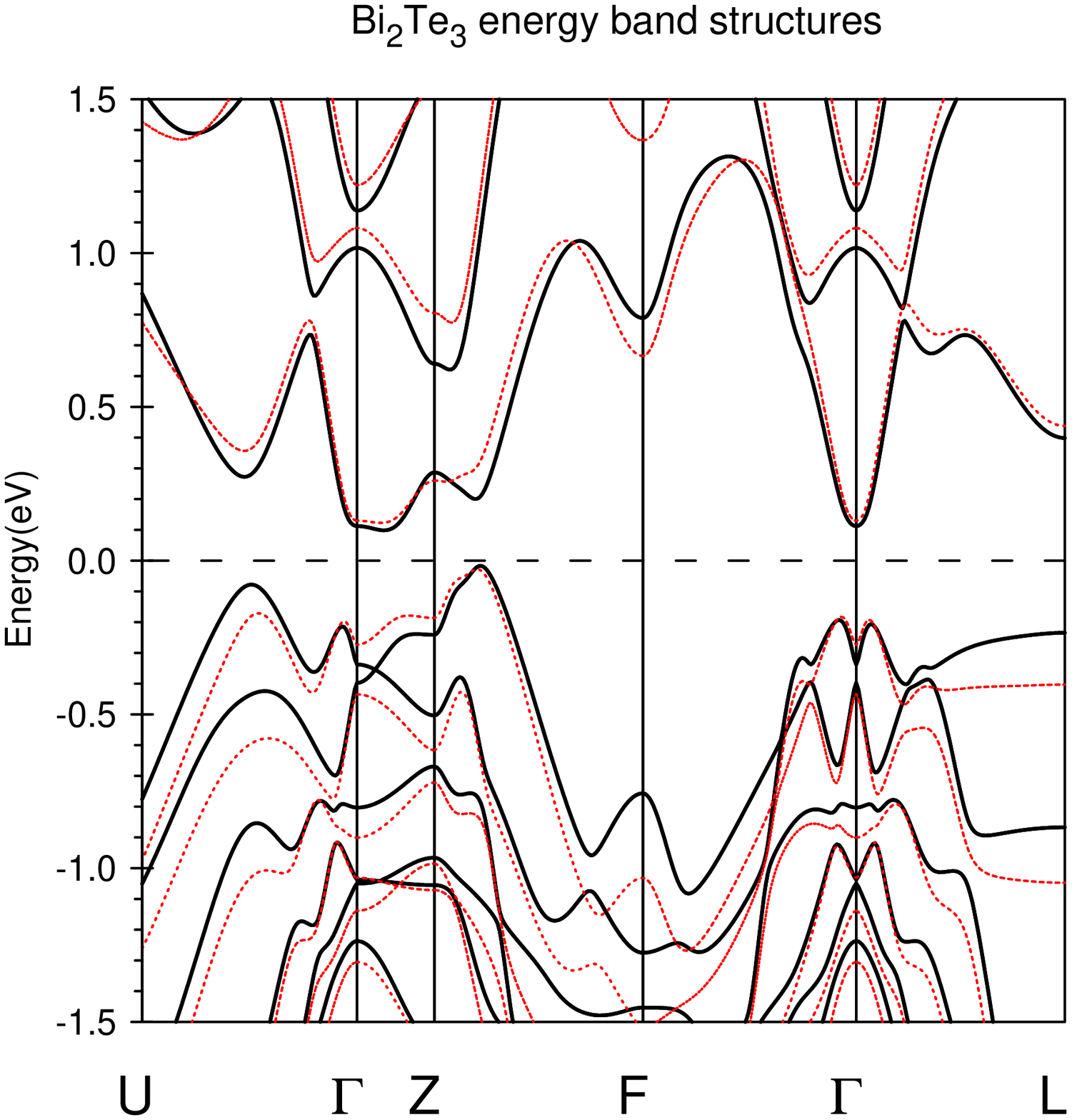}
\includegraphics[scale=0.35]{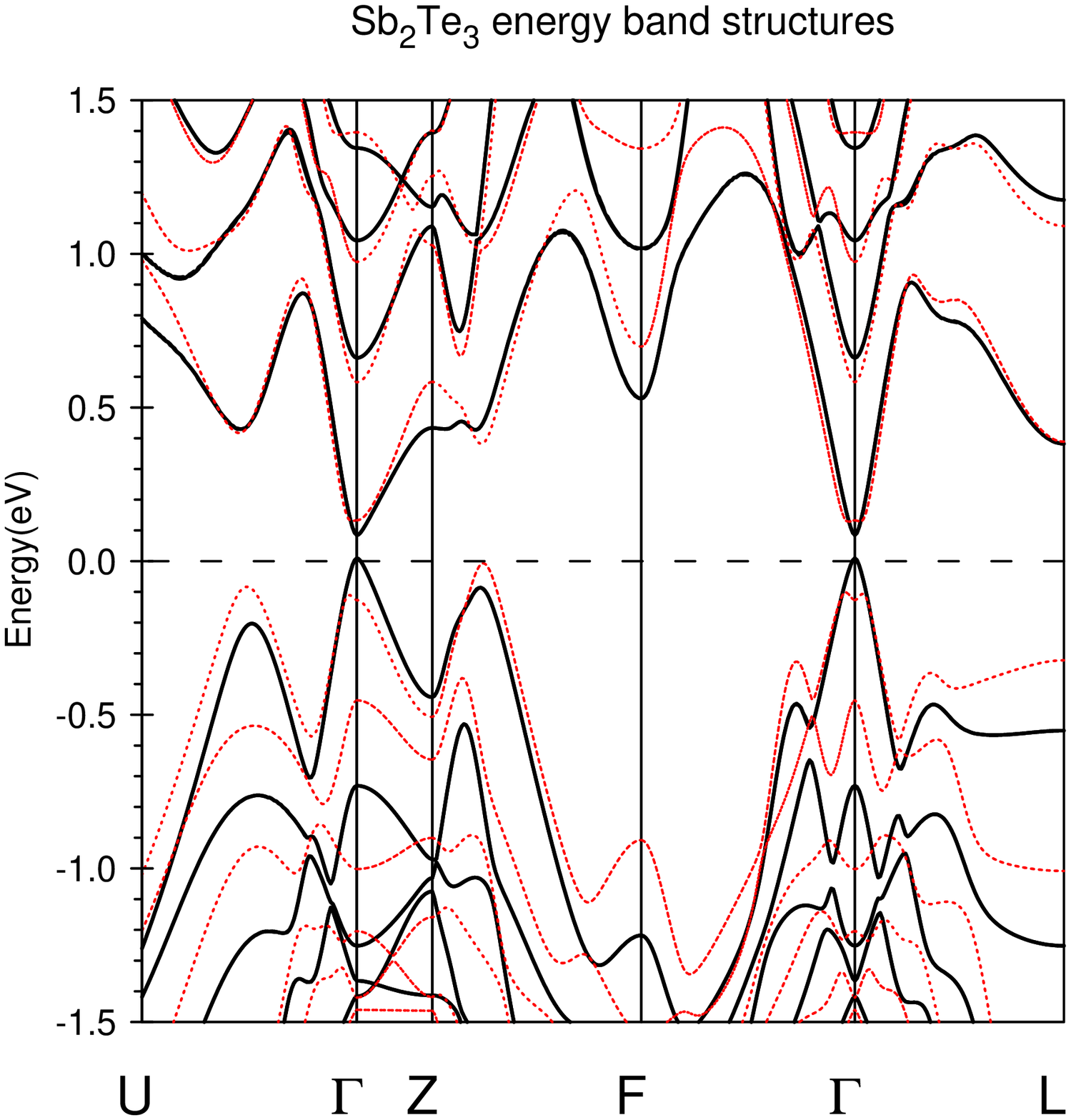}
\caption{\label{bands} Band structures of Bi$_2$Te$_3$ and Sb$_2$Te$_3$
along symmetry lines for both experimental (solid) and strained (dashed)
lattices. Energies are given relative to the VBM.}
\end{figure}

\begin{figure}
\includegraphics[scale=0.35]{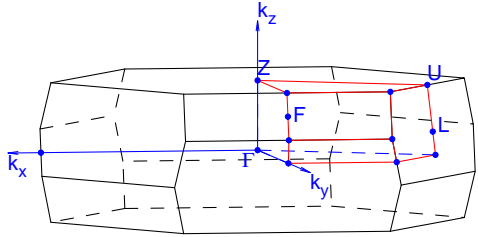}
\caption{\label{BZone} Brillouin zone of the rhombohedral lattice.}
\end{figure}

The electronic band structures for Bi$_2$Te$_3$ and Sb$_2$Te$_3$ for both
experimental and strained lattices are shown in Fig.\ref{bands}. 
The positions of the high-symmetry points in the BZ of the rhombohedral
structure are denoted in Fig.\ref{BZone}. Our results at the experimental
lattice parameters agree well with the previous {\it ab initio} studies of
Mishra {\it et al.}\cite{Mishra}, Larson {\it et al.}\cite{Larson1} for
Bi$_2$Te$_3$, and of Eremeev {\it et al.} \cite{Eremeev} both for Bi$_2$Te$_3$
and Sb$_2$Te$_3$. At the same time the calculations made with the
full-potential linearized augmented plane wave method (FLAPW) result in
slightly different band structures for the bismuth and antimony tellurides at
both the experimental\cite{Youn,Scheidemantel,Thonhauser1,Larson2,Huang} and 
strained \cite{Park,Wang} lattices. 
The key question of the band structure of Bi$_2$Te$_3$ and Sb$_2$Te$_3$ is the
position of the valence band maximum (VBM) and conduction band minimum (CBM)
in the BZ. The calculations of
Ref.[\onlinecite{Youn,Scheidemantel,Thonhauser1,Larson2,Huang}] result in a 
six-valley both VBM and CBM located in the symmetry plane ($\Gamma$ZU) in
agreement with experiments for both the bismuth\cite{bite6v} and
antimony\cite{sbte6v} tellurides. Unlike these results, in our case the CBM of 
Bi$_2$Te$_3$, at both lattice parameters, $a_{BiTe}$ and $a_{SbTe}$, is a
two-valley minimum located on the symmetry line $\Gamma$Z, similarly to
Ref.[\onlinecite{Mishra}]. For Sb$_2$Te$_3$ at the experimental lattice
parameters we found a direct band gap located at the center of the BZ, 
while at the larger in-plane lattice constant both six-valley VBM and CBM
lie in the symmetry plane ($\Gamma$ZU). In contrast to our results
Thonhauser\cite{Thonhauser2} found that the increase of the lattice parameters
in Sb$_2$Te$_3$ led to the formation of a direct band gap at the $\Gamma$ point.
On the other hand, the negative hydrostatic pressure discussed in
Ref.[\onlinecite{Thonhauser2}] implies an increase of both in-plane and
out-of-plane lattice parameters in comparable degree, while in our study
essentially the first one is included. Additionally, calculations in
Ref.[\onlinecite{Thonhauser2}] were performed for optimized atomic positions
with respect to the total energy, which can affect the band structure of
antimony telluride\cite{Wang}. 

As already discussed \cite{Mishra,Youn}, these differences in the band
structures are probably due to the non-spherical part of the potential, which
is not included in the ASA. At the same time, as discussed below, these
differences have no significant impact on the transport distribution $\sigma(E)$. The
details of the band structures for all four systems are compiled in Table I. 
\begin{table}
\caption{Band structure parameters: Band gap in eV, positions of VBM and CBM in
  crystallographic coordinates, effective masses in electron mass
  units, principal axes {\bf e}$_i$ in cartesian coordinates, and transport
  anisotropy ratio from the effective mass approximation.}
	\centering
		\begin{tabular}{|c|c|c|c|c|}
		  \hline
		\multicolumn{5}{|c|}{Bi$_2$Te$_3$}\\ \hline
	       Lattice parameters&\multicolumn{2}{|c|}{$a=a_{BiTe}$}&\multicolumn{2}{|c|}{$a=a_{SbTe}$}\\ \hline
		Gap(eV)&\multicolumn{2}{|c|}{0.105}&\multicolumn{2}{|c|}{0.129}\\ \hline
		Extremum         & VBM                 & CBM                 & VBM                 & CBM\\ \hline
		Position         &  0.517  0.366  0.366&  0.173  0.173  0.173&  0.405  0.405  0.335&  0.151  0.151  0.151\\ \hline
	        Effective masses &                     &                     &                     &\\
                $m_1$            & -0.024              &  0.178              & -0.039              & 0.154\\
                $m_2$            & -0.134              &  0.178              & -0.077              & 0.154\\
                $m_3$            & -1.921              &  0.835              & -0.207              & 1.370\\\hline
		Principal axes   &                     &                     &                     & \\
                {\bf e}$_1$      &  0.500 -0.867  0.000&  1.000  0.000  0.000&  0.866  0.499 -0.024&  1.000  0.000  0.000\\
                {\bf e}$_2$      &  0.600  0.346  0.721&  0.000  1.000  0.000&  0.500 -0.867  0.000&  0.000  1.000  0.000\\
                {\bf e}$_3$      &  0.625  0.361 -0.693&  0.000  0.000  1.000&  0.021  0.012  0.999&  0.000  0.000  1.000\\\hline
       $\sigma_{xx}/\sigma_{zz}$ &  5.452              &  4.700              &       4.020         &   9.013\\ \hline
		\multicolumn{5}{|c|}{Sb$_2$Te$_3$}\\ \hline
		Latt. param. &\multicolumn{2}{|c|}{$a=a_{SbTe}$}&\multicolumn{2}{|c|}{$a=a_{BiTe}$}\\ \hline
		Gap(eV)&\multicolumn{2}{|c|}{0.090}&\multicolumn{2}{|c|}{0.140}\\ \hline
		Extremum         & VBM                 & CBM                 & VBM                 & CBM\\ \hline
		Position         &  0.000  0.000  0.000&  0.000  0.000  0.000&  0.547  0.392  0.392&  0.004  0.020  0.020  \\ \hline
		Effective masses &                     &                     &                     &\\
                $m_1$            & -0.054              &  0.045              &  -0.039             &   1.124               \\
                $m_2$            & -0.054              &  0.045              &  -0.083             &   1.774               \\
                $m_3$            & -0.102              &  0.114              &  -2.046             &   6.861               \\\hline
		Principal axes   &                     &                     &                     &\\
                {\bf e}$_1$      &  1.000  0.000  0.000&  1.000  0.000  0.000&  0.500 -0.867  0.000&  -0.316 -0.183  0.931\\
                {\bf e}$_2$      &  0.000  1.000  0.000&  0.000  1.000  0.000&  0.594  0.343  0.727&   0.500 -0.867  0.000\\
                {\bf e}$_3$      &  0.000  0.000  1.000&  0.000  0.000  1.000&  0.630  0.363 -0.686&   0.806  0.465  0.365\\\hline
       $\sigma_{xx}/\sigma_{zz}$ &  1.889              &  2.507              &  2.397              &   2.080              \\ \hline

		\end{tabular}
\end{table}

\begin{figure}
\includegraphics[scale=0.40]{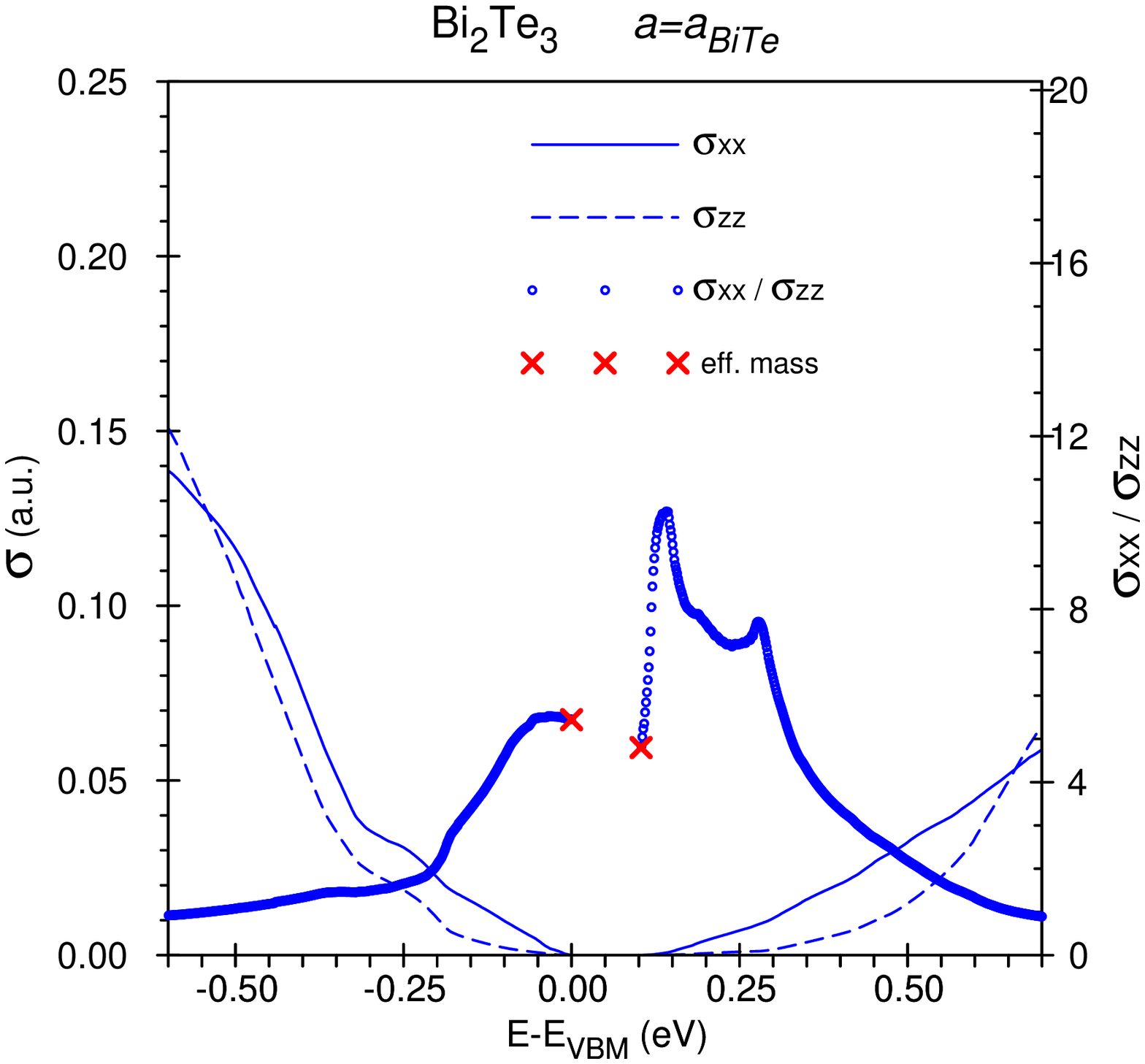}
\includegraphics[scale=0.40]{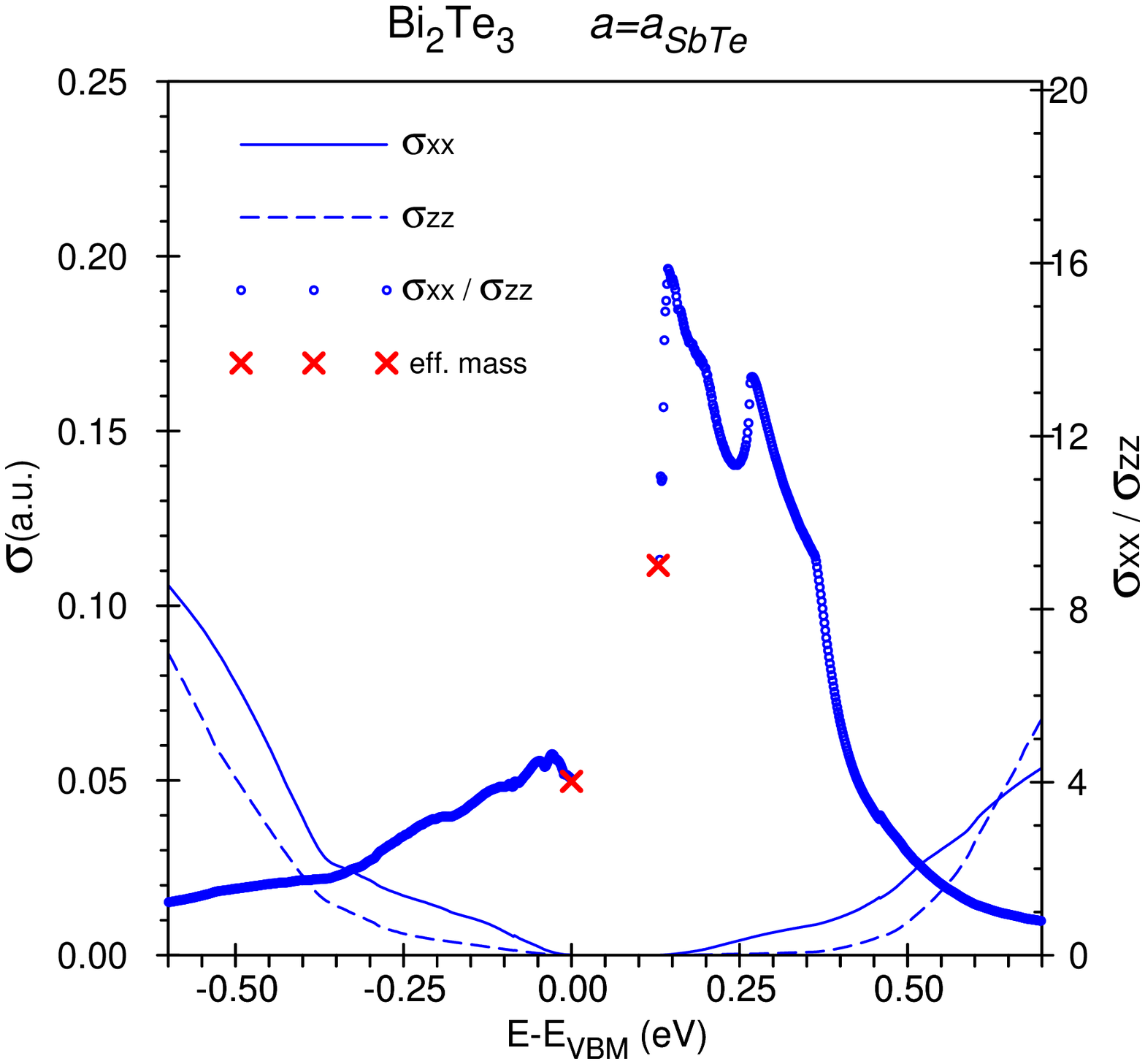}
\caption{\label{condbite} Transport distribution $\sigma_{xx}(E)$ and
  $\sigma_{zz}(E)$ and the transport anisotropy $\sigma_{xx}/\sigma_{zz}$ for
  Bi$_2$Te$_3$ at the experimental and compressed lattice parameters. Crosses
  at the band edges mark the $\sigma_{xx}/\sigma_{zz}$ ratio derived from the 
  effective mass model \cite{Peter} using the parameters of Table I. }
\end{figure}

\begin{figure}
\includegraphics[scale=0.40]{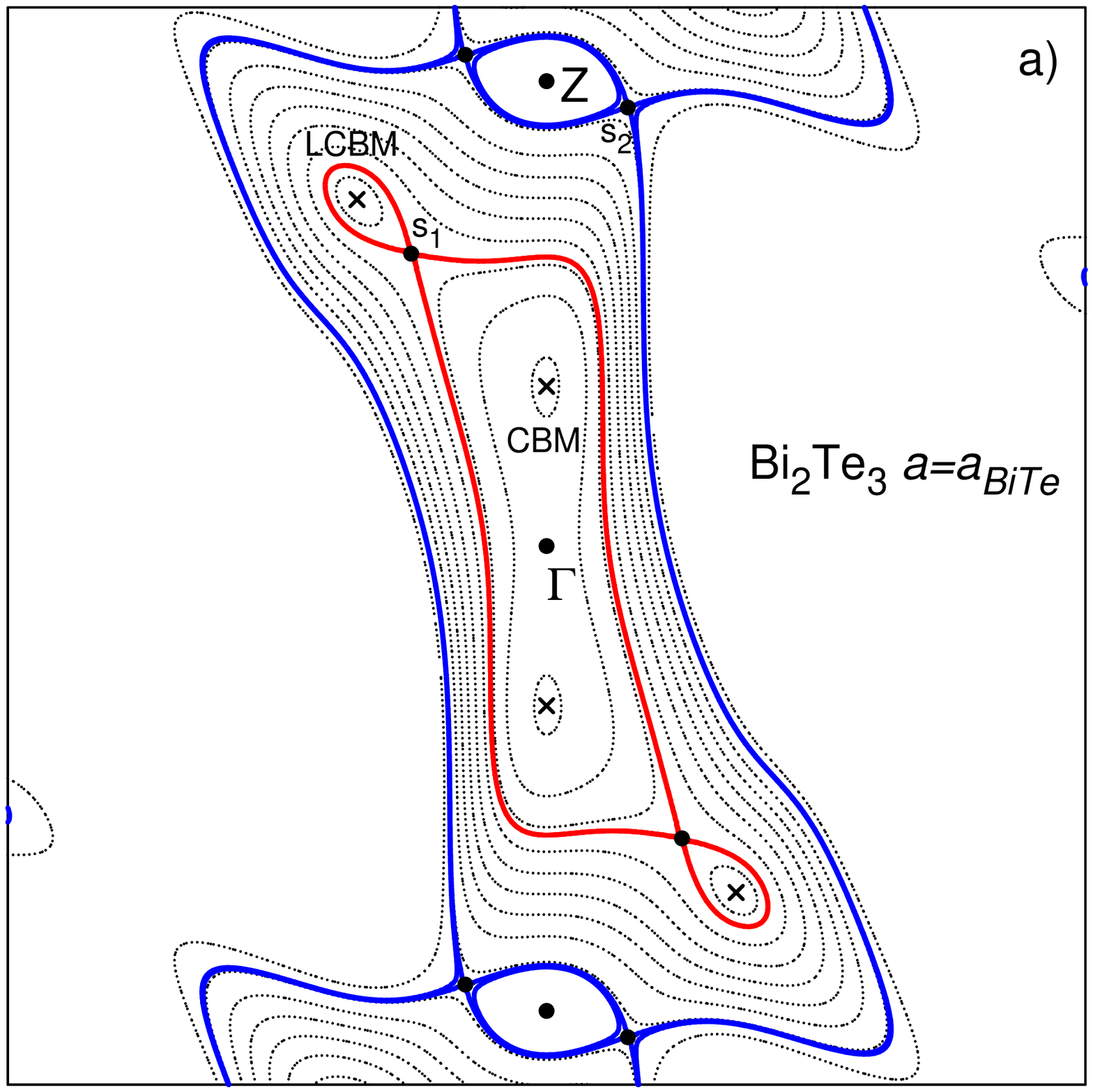}
\includegraphics[scale=0.40]{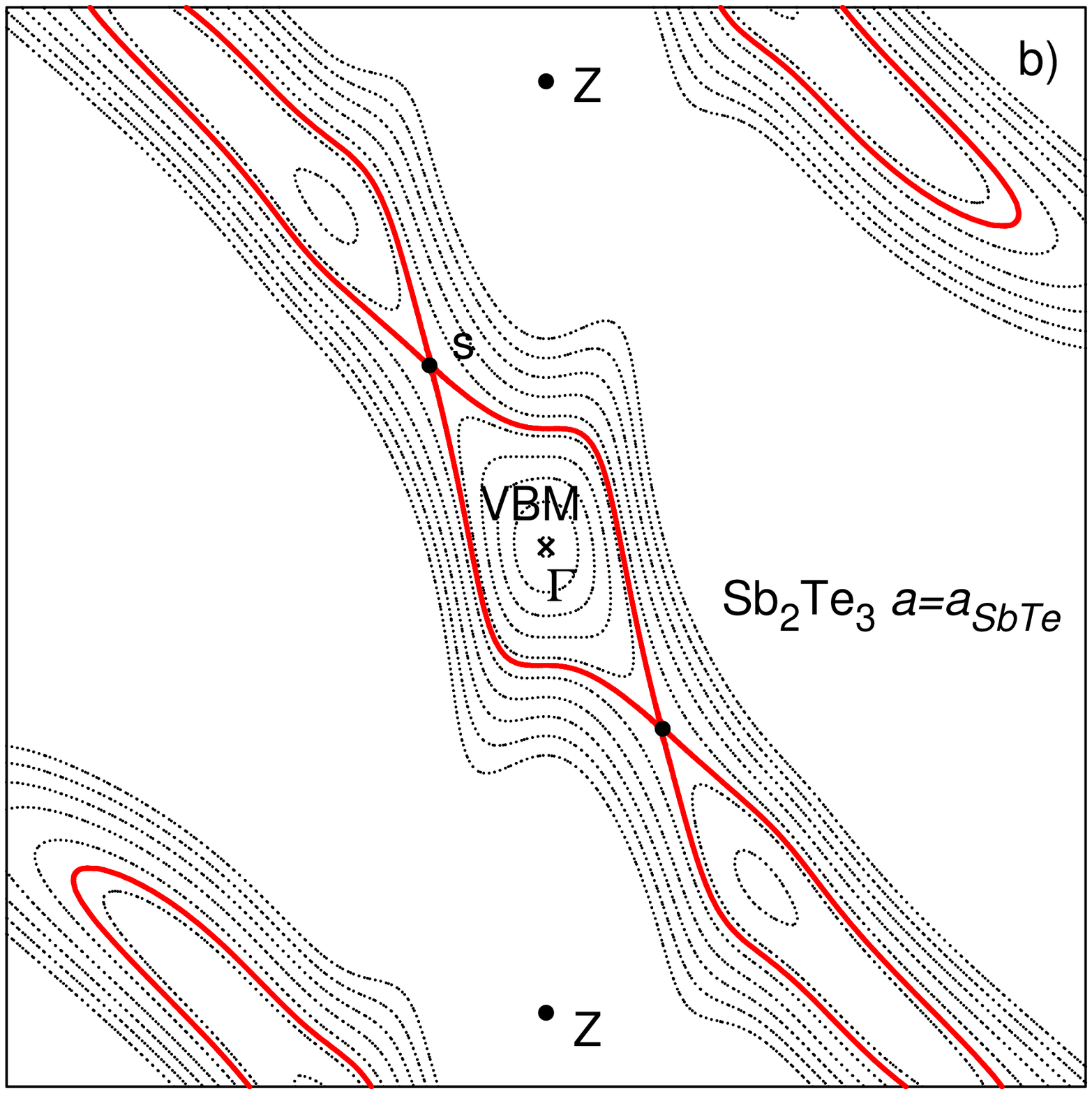}
\caption{\label{iso} Contour plots of $\varepsilon ({\bf k})$ at the
  experimental lattice constants in the plane
  ($\Gamma$ZU)\\(a) Bi$_2$Te$_3$, 10 isolines for ($E-E_{CBM}$) at 0 to 0.19eV
  with a constant increment(dotted), additionally, 2 isolines at
  $E-E_{CBM}=0.04$eV and $E-E_{CBM}=0.17$eV with the saddle points $s_1$ and
  $s_2$, respectively (bold), the positions of the CBM and the LCBM
  are marked with crosses.\\(b) Sb$_2$Te$_3$, 10 isolines for ($E-E_{VBM}$) at
  -0.24 eV to 0 with a constant increment (dotted), additionally, 1 isoline
  with the saddle point $s$ (bold), the position of the VBM is
  marked with a cross.}
\end{figure}

The calculated transport distribution of Bi$_2$Te$_3$ at both the
experimental, $a=a_{BiTe}$, and compressed, $a=a_{SbTe}$, lattice parameters
are shown in Fig.\ref{condbite} together with the anisotropy ratio
$\sigma_{xx}/\sigma_{zz}$. In terms of the rigid band model the energies below
and above the band gap simulate $p$- and $n$-doping respectively. While
for $p$-doping close the VBM $\sigma_{xx}/\sigma_{zz}$ varies
smoothly approaching the limiting value, the ratio increases drastically and
forms a prominent two-peak structure  for $n$-doping case.
This structure originates from the two topological transformations of the
constant energy surfaces in the conduction band. Fig.\ref{iso}(a) shows the
contour plot of $\varepsilon ({\bf k})$ for Bi$_2$Te$_3$ at $a=a_{BiTe}$ in
the plane ($\Gamma$ZU) for energies 0 to 0.19eV relative to the conduction band
edge. The main features of the band structure  are the global CBM on the line $\Gamma$Z,
the local conduction band minimum (LCBM) at (0.666,0.602,0.602), and two
saddle points $s_1$ at (0.722 ,0.667, 0.667) and $s_2$ at
(0.493,0.461,0.461), in crystallographic coordinates. The saddle point $s_1$
occurs at $E-E_{CBM}=0.04$eV and causes the first peak of
$\sigma_{xx}/\sigma_{zz}$, while $s_2$ appears at 0.17eV and forms the second peak. 
At room temperature the chemical potential would be located in the saddle point $s_1$ or $s_2$, 
for an electron carrier concentration of about $N = 3.0 \times 10^{19}{cm^{-3}}$ or 
$N = 1.5 \times 10^{20}{cm^{-3}}$, respectively.
The location of the LCBM in our band structure calculation is close to the
position of the six-valley CBM reported in
Ref.[\onlinecite{Youn,Scheidemantel,Thonhauser1,Larson2,Huang,Wang}], and the
energy difference between these two minima,$~E_{LCBM}-E_{CBM}$=0.04eV, is
quite small. At the same time the saddle point $s_2$ close to the local
band maximum at Z was found  as well in agreement with these calculations. A
slight shift of the energies at the CBM and the LCBM would bring these two
band structures in accordance. Since the saddle point $s_1$ lies close to the
line connecting the CBM and the LCBM this modification would not affect
remarkably the band structure topology. Moreover, the transport anisotropy
$\sigma_{xx}/\sigma_{zz}$ at the LCBM from the effective mass approximation is
4.95, which is fairly close to 4.7 at the CBM. This indicates that the
transport anisotropy of the bismuth telluride at the experimental lattice
parameters is stable with respect to small modifications of the band
structures with a two- and six-valley CBM, respectively.

In Bi$_2$Te$_3$ the in-plane compression of the lattice parameters from
$a=a_{BiTe}$ to $a=a_{SbTe}$ increases the transport anisotropy ratio 
at the conduction band edge remarkably. Within the effective mass
approximation this can be explained by the enhancement of the ratio
$m_{\perp}/m_{\parallel}$ due to the expansion of the BZ in $xy$ plane. At the
same time $\sigma_{xx}/\sigma_{zz}$ decreases at the valence band edge. 
In this case the compression of the lattice results in a re-orientation of the
longest axis of the effective mass ellipsoid near the VBM closer to the $z$
axis, which lead to an increase of the transport anisotropy ratio, and at the
same time decreases the anisotropy of the dominating effective masses $m_2$
and $m_3$ (see Table I). The last effect prevail over the enhancement of
$\sigma_{xx}/\sigma_{zz}$ due to the rotation of the effective mass ellipsoid.
\begin{figure}
\includegraphics[scale=0.40]{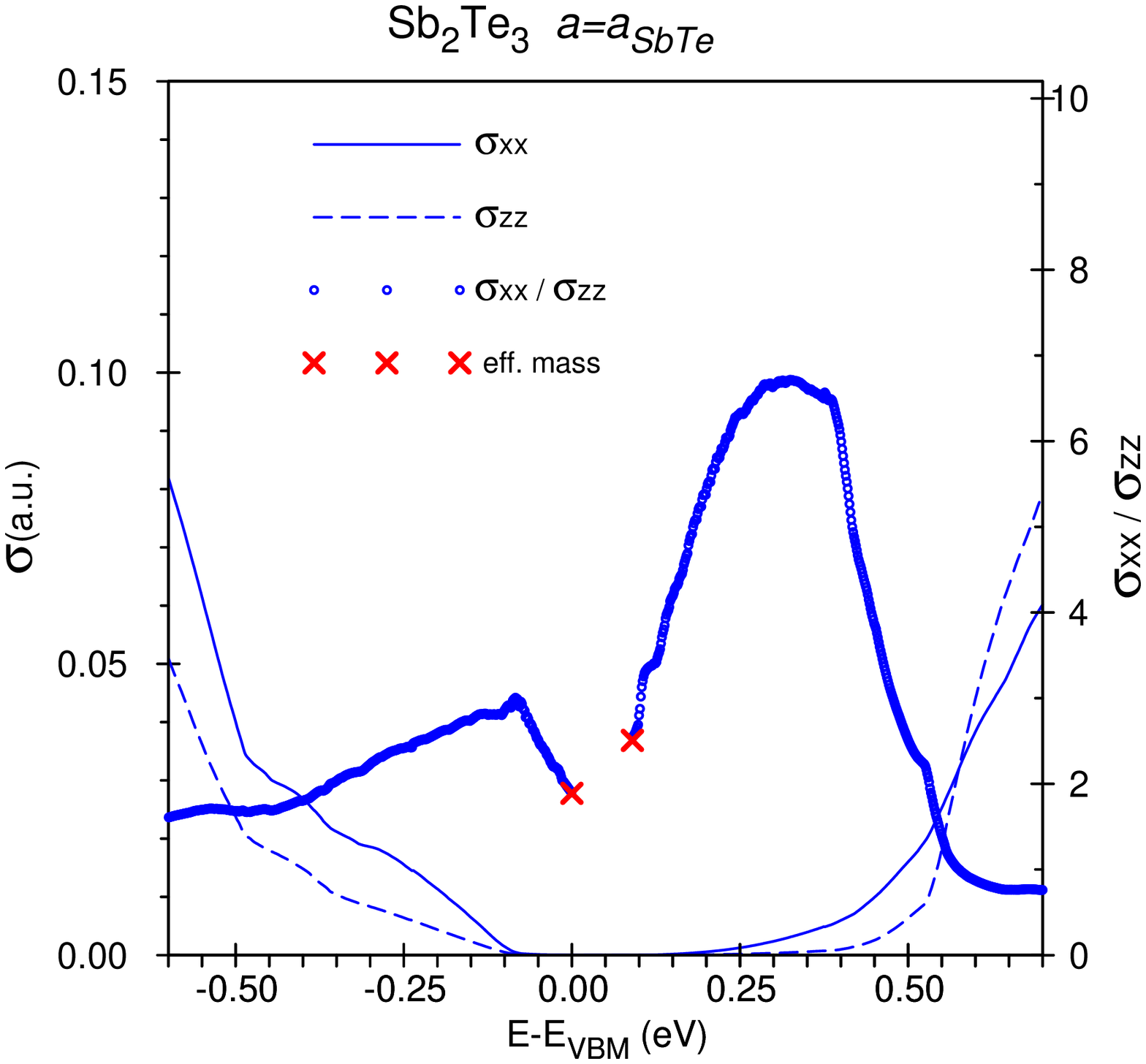}
\includegraphics[scale=0.40]{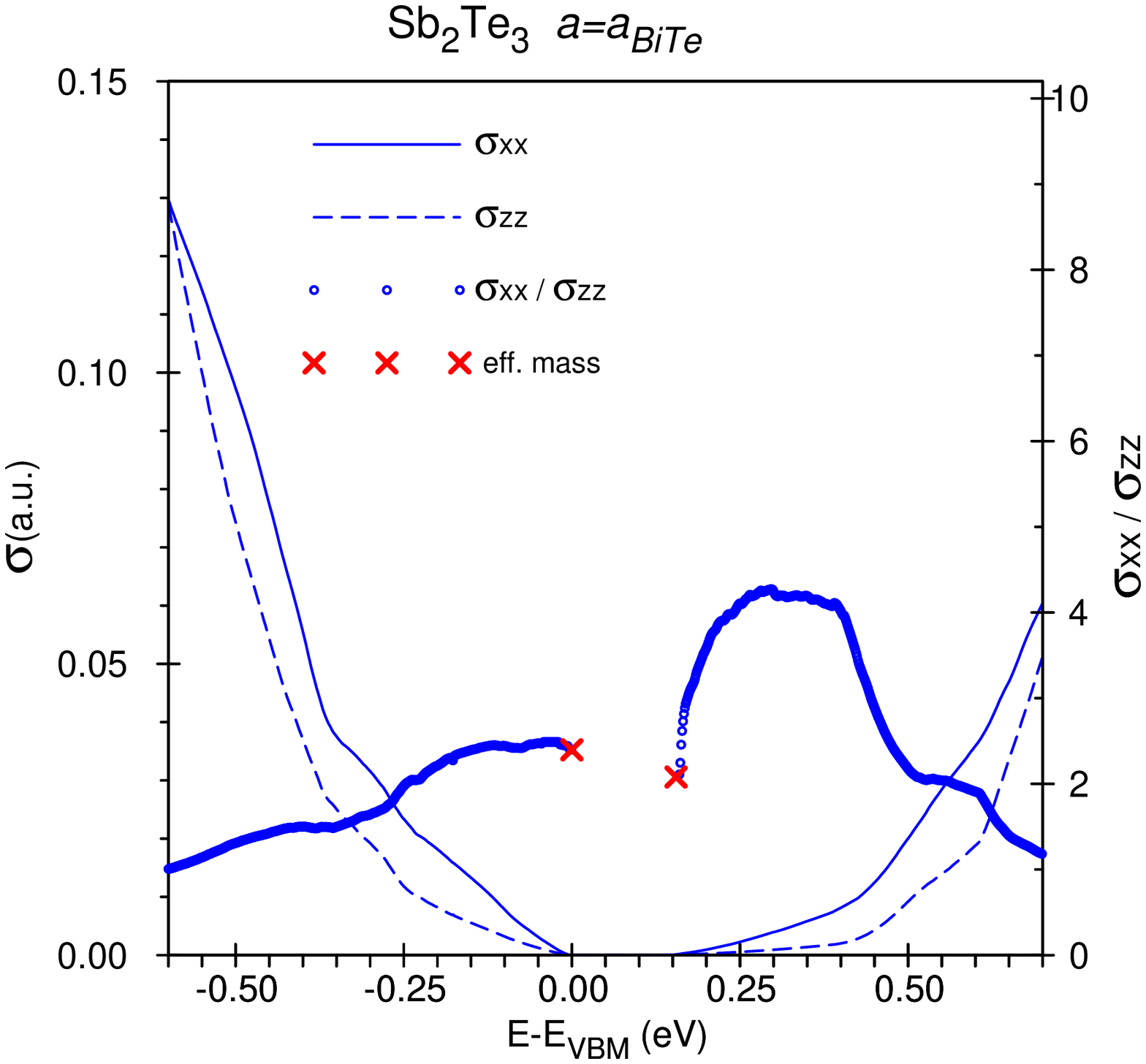}
\caption{\label{condsbte} Transport distribution and the transport anisotropy
  for Sb$_2$Te$_3$ at the experimental and expanded lattice parameters. Crosses
  at the band edges mark the $\sigma_{xx}/\sigma_{zz}$ derived from the
  effective mass model \cite{Peter} using the parameters of Table I.}
\end{figure}
Fig.\ref{condsbte} shows the transport distribution and the anisotropy ratio
of Sb$_2$Te$_3$ at both $a=a_{SbTe}$ and $a=a_{BiTe}$. The kink of the
anisotropy ratio at the experimental lattice parameters is induced by
the saddle point $s=$(0.831, 0.784, 0.784) at $E-E_{VBM}=0.116$ eV, which 
corresponds to a hole carrier concentration of $N = 5.8 \times 10^{19}{cm^{-3}}$ 
at 300K. This topology is illustrated in Fig.\ref{iso}(b), which shows the
contour plot of $\varepsilon ({\bf k})$ in the valence band of the antimony
telluride for energies -0.24 eV to 0 relative to $E_{VBM}$. The in-plane
expansion of the lattice parameter increases the density of the occupied
states near the valence band edge and suppresses the kink. The transport
anisotropy ratio increases with the in-plane compression of the BZ at the
valence band edge, and, at the same time, decreases at the conduction band
edge due to the larger angle between the $z$ axis and the largest axis of the
effective mass ellipsoid, similarly to the discussed changes in bismuth
telluride.
\section{Conclusions}
On the basis of {\it ab initio} electronic structures obtained with the fully
relativistic KKR method we studied the anisotropy of the transport properties
of the bismuth and antimony tellurides in the constant relaxation time
approximation within the Boltzmann formalism. In addition to the systems with
the experimental lattice parameters we modeled bismuth telluride within the
lattice of Sb$_2$Te$_3$, and vice versa. We found that a decrease of
the in-plane lattice parameters increases the transport anisotropy for the
$n$-doping and, at the same time, decreases the anisotropy for the $p$-doped
case. This effect can be understood within the effective mass approximation at
the valence band maximum and conduction band minimum, respectively.

%\begin{acknowledgments}
%...
%\end{acknowledgments}
\newpage %Just because of unusual number of tables stacked at end
%\bibliography{paper}% Produces the bibliography via BibTeX.

\begin{thebibliography}{99}
\bibitem[*]{byline} Corresponding author.\\
Email address: {\texttt bogdan.yavorsky@physik.uni-halle.de}

\bibitem{DiSalvo}
F. J. DiSalvo, Science {\bf285}, 703 (1999) 

\bibitem{Bell}
L. E. Bell, Science {\bf321}, 1457 (2008) 

\bibitem{exp_early}
C. B. Satterthwaite and R. W. Ure,  Phys. Rev. {\bf108}, 1164 (1957),\\
F. D. Rosi, B. Abeles, and R. S. Jensen,  J. Phys. Chem. Solids {\bf10}, 191
(1959)
 
\bibitem{exp_further}
J. P. Fleurial, L. Gailliard, R. Triboulet, H. Scherrer, and S. Scherrer,
J. Phys. Chem. Solids {\bf49}, 1237 (1988),\\ 
T. Caillat, M. Carle, P. Pierrat, H. Scherrer, and S. Scherrer,
J. Phys. Chem. Solids {\bf53}, 1121 (1992) 

\bibitem{Rama1}
R. Venkatasubramanian, T. Colpitts, B. O'Quinn, S. Liu, , N. El-Masry, and
M. Lamvik, Appl.Phys.Lett. {\bf75}, 1104 (1999)

\bibitem{Rama2}
R. Venkatasubramanian, E. Siilova, T. Colpitts, and B. O'Quinn, Nature
{\bf413}, 597 (2001)

\bibitem{Thomas}
G. A. Thomas, D. H. Rapkine, R. B. Van Dover, L. F. Mettheiss, W. A. Sunder,
L. F. Schneemeyer, and J. V. Waszczak, Phys. Rev. B {\bf46}, 1553 (1992)

\bibitem{Mishra}
S. K. Mishra, S. Satpathy, and O. Jepsen, J.Phys.: Condens. Matter {\bf9}, 461 (1997)

\bibitem{Larson1}
P. Larson, S. D. Mahanti, and M. G. Kanatzidis,  Phys. Rev. B {\bf61}, 8162 (2000)

\bibitem{Youn}
S. J. Youn and A. J. Freeman, Phys. Rev. B {\bf63}, 085112 (2001)

\bibitem{Scheidemantel}
 T. J. Scheidemantel, C. Ambrosch-Draxl, T. Thonhauser, J. V. Badding,
 and J. O. Sofo, Phys. Rev. B {\bf68},125210 (2003)

\bibitem{Thonhauser1}
T. Thonhauser, T. J. Scheidemantel, J. O. Sofo, J. V. Badding, and G. D. Mahan,
 Phys. Rev. B {\bf68},085201 (2003)

\bibitem{Larson2}
P. Larson, Phys. Rev. B {\bf68}, 155121 (2003)

\bibitem{Thonhauser2}
T. Thonhauser, Solid State Commun. {\bf129}, 249 (2004)

\bibitem{Kim}
M. Kim, A. J. Freeman, and C. B. Geller,  Phys. Rev. B {\bf72}, 035205 (2005)

\bibitem{Larson3}
P. Larson, Phys. Rev. B {\bf74}, 205113 (2006)

\bibitem{Lee}
S. Lee, and P. von Allmenn,  Appl. Phys. Lett. {\bf88}, 022107 (2006)

\bibitem{Wang}
G. Wang and T. Cagin, Phys. Rev. B {\bf76}, 075201 (2007)


\bibitem{Huang}
B.-L. Huang, and M. Kaviany, Phys. Rev. B {\bf77}, 125209 (2008)

\bibitem{Park}
M. S. Park, J.-H. Song, J. E. Medvedeva, M. Kim, I. G. Kim, and A. J, Freeman,
Phys. Rev. B {\bf81}, 155211 (2010)

\bibitem{Eremeev}
S. V. Eremeev, Yu. M. Koroteev, and E. V. Chulkov, JETP Letters {\bf91}, 387 (2010)

\bibitem{Li}
H. Li, D. Bilc, and S. D. Mahanti, Mat. Res. Soc. Symp. Proc. {\bf793}, 8.37 (2004)

\bibitem{SpringerData}
O. Madelung, M. Schulz, H. Weiss (Eds.) Landolt-B{\"o}rnstein, New Series,
Group III, vol. 17f, (Springer, New York, 1983) \\
R. W. G. Wyckoff , Crystal Structures 2, J. (Wiley and Sons, New York, 1964)\\
Th. L. Anderson , H. Krause, H. Brigitte: Acta Crystallogr. B 30, 1307 (1974)

\bibitem{KKR}
R. Zeller, P. H. Dederichs, B. {\'U}jfalussy, L. Szunyogh, and. P. Weinberger,
Phys. Rev. B {\bf52}, 8807 (1995)\\
N. Papanikolau, R. Zeller, and P. H. Dederichs, J. Phys.: Condens. Matter
{\bf14}, 2799 (2002)

\bibitem{Vosko}
S. H. Vosko, L. Wilk, and M. Nusair, Can. J. Phys. {\bf58}, 1200 (1980)

\bibitem{KKRrel}
M. Gradhand, M. Czerner, D. V. Fedorov, P. Zahn, B. Yu. Yavorsky, L. Szunyogh,
and I. Mertig, Phys. Rev. B {\bf80}, 224413 (2009)

\bibitem{Ziman} J. M. Ziman, {\it Principles of the Theory of Solids}
  (Cambridge University Press, Cambridge, 1972).  

\bibitem{Peter} P. Zahn, N. F. Hinsche, B. Yu. Yavorsky, and I. Mertig, to be
  published (2011)

\bibitem{bite6v}
J. R. Drabble, R. D. Groves, and R. Wolfe, Proc. Phys. Soc. London {\bf71}, 430 (1958)\\
R. B. Mallinson, J. A. Rayne, and R. W. Ure, Jr., Phys. Lett {\bf19}, 545 (1965)
\bibitem{sbte6v}
H. Schwarz, G. Bj{\"o}rck, and O. Beckman , Soli. State Commun. {\bf5}, 905 (1967)\\
V. A. Kulbachinskii, Z. M. Dashevskii, M. Inoue, M. Sasaki, H. Negishi, W. X. Gao, 
P. Lostak, J. Horak, and A. de Visser, Phys. Rev. B {\bf52}, 10915 (1995)

\bibitem{Bloechl}
P. E. Bl{\"o}chl, O. Jepsen, and O. K. Andersen,  Phys. Rev. B {\bf49}, 16223 (1994) 
\end{thebibliography}

\end{document}